\documentclass[twocolumn]{aastex63}
\usepackage[utf8]{inputenc}
\usepackage{amsmath,amsfonts,amssymb}
\usepackage{graphicx,subfigure,multirow}
\usepackage{epsfig,color}
\usepackage{enumitem}
\usepackage{natbib}
\usepackage{hyperref}
\usepackage{verbatim}
\usepackage{lipsum}
\usepackage{longtable}
\usepackage{booktabs}

\begin{document}
\title{Bayesian model-selection of neutron star  equation of state using multi-messenger observations}

\author[0000-0003-2131-1476]{Bhaskar Biswas}
\affiliation{Inter-University Centre for Astronomy and Astrophysics, Post Bag 4, Ganeshkhind, Pune 411 007, India}
\affiliation{The Oskar Klein Centre, Department of Astronomy, Stockholm University, AlbaNova, SE-10691 Stockholm,
Sweden}

\begin{abstract}
Measurement of macroscopic properties of neutron stars, whether in binary or in an isolated system, provides us a key opportunity to place a stringent constraint on its equation of state. In this paper, we perform Bayesian model-selection on a wide variety of neutron star equation of state using multi-messenger observations. In particular, (i) we use the mass and tidal deformability measurement from two binary neutron star merger event, GW170817 and GW190425; (ii) simultaneous mass-radius measurement of PSR J0030+0451 and PSR J0740+6620 by NICER collaboration, while the latter has been analyzed by joint NICER/radio/XMM-Newton collaboration. Among the 31 equations of state considered in this analysis, we are able to rule out different variants of MS1 family, SKI5, H4, and WFF1 EoSs decisively, which are either extremely stiff or soft equations of state. The most preferred equation of state model turns out to be AP3 (or MPA1), which predicts the radius and dimensionless tidal deformability of a $1.4 M_{\odot}$ neutron star to be 12.10 (12.50) km and 393 (513) respectively.
\end{abstract}

\section{Introduction} Neutron stars (NSs) are the extremely dense objects known in our universe. Properties of matter inside NS are encoded in its equation of state (EoS), which has wide-ranging uncertainty from the theoretical perspective~\citep{Lattimer_2016,Oertel_2017,Baym_2018}. With the current understanding of quantum chromodynamics, it is very hard to determine the interactions of NS matter at such high densities. Also performing many-body calculations is computationally intractable. Besides, the constituent of NS at its core is highly speculative -- perhaps containing exotic matter like quark matter, strange baryons, meson condensates etc.~\citep{Glendenning:1997wn}. Even though the matter inside the NS is extremely dense, nevertheless the temperature of this object is actually cold for most of its life span. Such highly dense but rather cold matter cannot be produced in the laboratory. Since probing the physics of NS matter is inaccessible by our earth based experiments, we look for astrophysical observations of NS. These observations can give us the measurement of macroscopic properties related to a NS, like mass, radius, tidal deformability~\citep{Hinderer:2007mb,Binnington:2009bb,Damour:2009vw} etc., which will depend on the internal structure of the NS. The measurement of these macroscopic properties can be used to infer the EoS of NS.

The past few years have been a golden era in NS physics. A number of key astrophysical observations of NS have been made from multiple cosmic messengers such as radio observations of massive pulsars~\citep{Demorest:2010bx,Antoniadis:2013pzd,Cromartie:2019kug,2021arXiv210400880F} or mass-radius measurement of PSR J0030+0451~\citep{Miller:2019cac,Riley:2019yda} and PSR J0740+6620~\citep{Miller:2021qha,Riley:2021pdl} using pulse-profile modeling by NICER collaboration~\citep{2016SPIE.9905E..1HG}, and also the observation of gravitational waves (GW) from two coalescing binary neutron star (BNS) merger  events~\citep{TheLIGOScientific:2017qsa,Abbott:2018exr,Abbott:2020uma} by LIGO/Virgo collaboration~\citep{advanced-ligo,advanced-virgo}. These observations have motivated several authors to place joint constraints on the properties of NS~\citep{Raaijmakers:2019dks,Capano:2019eae,Landry_2020PhRvD.101l3007L,Jiang:2019rcw,Traversi:2020aaa,Biswas_arXiv_2008.01582B,Al-Mamun:2020vzu,Dietrich:2020efo,Biswas:2021yge,Breschi_2021,Miller:2021qha,Raaijmakers:2021uju,Pang:2021jta} either using a phenomenological or nuclear-physics motivated EoS parameterization by employing Bayesian parameter estimation. However, one can take a complementary approach instead of constructing an EoS parameterization and ask the following question: given a variety of NS EoS models in the literature based on nuclear-physics, which one is the most preferred by the current observations in a statistical sense?  

A few studies~\citep{LIGOScientific:2019eut,Ghosh:2021eqv,Pacilio:2021jmq} already exist in the literature on Bayesian model-selection of NS EoS. But they only use GW observations and do not provide the current status of various NS EoS models. Therefore, the aim of this paper is to perform a Bayesian model-selection study amongst various nuclear-physics motivated EoS models of NS using the constraints coming from multi-messenger astronomy.
\begin{figure}[ht]
    \centering
    \includegraphics[width=0.45\textwidth]{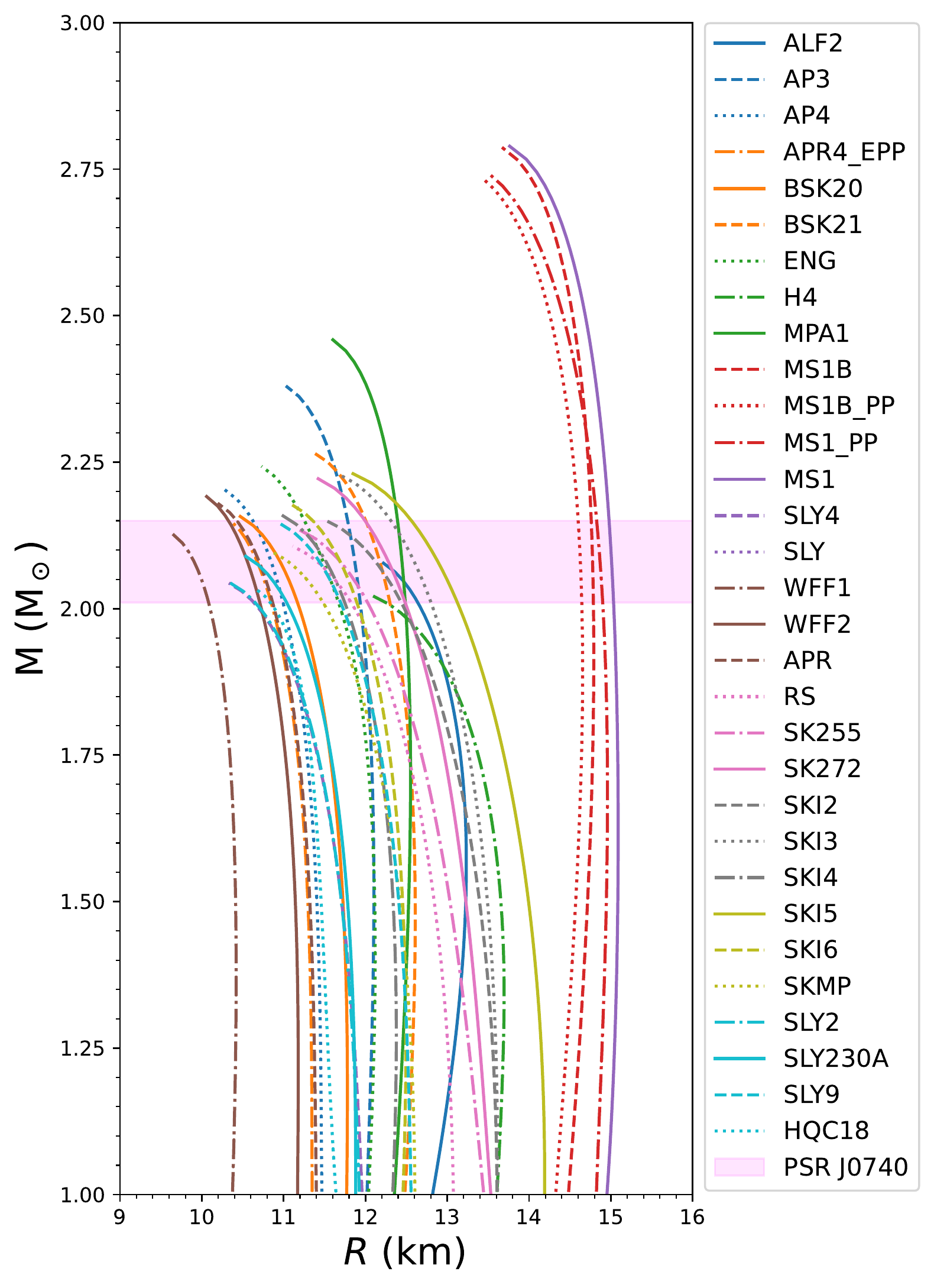}
    \caption{The mass-radius ($M-R$) diagram for all the 31 EoSs are shown here. The magenta band corresponds to $1 \sigma$ confidence interval of the mass measurement of PSR J0740+6620.}
    \label{fig:MR_laleos}
\end{figure}

\section{EoS catalog} For this work, we consider 31 EoS models which are computed from different nuclear-physics approximations covering a wide-range in mass-radius (or equivalently pressure-density) diagram. We take these EoSs from publicly available~{\tt LalSuite}~\citep{lalsuite} package and also use their code to calculate all the relevant macroscopic properties such as mass, radius, and tidal deformability. Most of these EoSs are consisted of plain $npe\mu$ nuclear matter which include--- \\
(i) Variational-method EoSs (AP3-4) and APR~\citep{APR4_EFP}, APR4\_EFP~\citep{Endrizzi_2016,APR4_EFP}, WFF1-2~\citep{wff1}), (ii) potential based EoS SLY~\citep{Douchin:2001sv}, (iii) nonrelativistic Skyrme interactions based EoS (SLY2 and SLY9~\citep{SLy2_1,SLy2_2}, SLY230A~\citep{SLy230A_3}, RS~\citep{RS_3}, BSK20 and BSK21~\citep{Goriely_2010,Pearson_2011}, SK255 and SK272~\citep{SLy2_1,SLy2_2,SK255_3}, SKI2-6~\citep{SLy2_1,SLy2_2,REINHARD1995467}, SKMP~\citep{SLy2_1,SLy2_2,SKMP_3}), (iv) relativistic Brueckner-Hartree-Fock EOSs (MPA1~\citep{MPA1}, ENG~\citep{Engvik_1994}), (v) relativistic mean field theory EoSs (MS1, MS1B, MS1\_PP, MS1B\_PP where MS1\_PP, MS1B\_PP~\citep{MS_PP} are the analytic piecewise polytrope fits of original MS1 and MS1B EoS, respectively). Also we consider one model with hyperons H4~\citep{H4}, and nucleonic matter mixed with quark EoSs --- ALF2~\citep{Alf2} and HQC18~\citep{HQC18}. In Table~\ref{tab:eos_table}, the radius and tidal deformability of $1.4 M_{\odot}$ NS are shown for each EoS and they lie in the range (10.42, 15.07) km and (153,1622) respectively. This is to note that for the choice of EoS catalog we follow Ref.~\citep{LIGOScientific:2019eut} that excludes EoSs with phase transition except HQC18~\citep{HQC18}. Therefore, for the details on these EoSs, readers are referred to the respective references listed in Table~\ref{tab:eos_table} and as well as Ref.~\citep{LIGOScientific:2019eut}. These EoSs are also chosen in such a way that they are compatible with mass measurement ($M =2.08 \pm 0.07 M_{\odot}$ at $1 \sigma$ confidence interval) of the observed heaviest pulsar~\citep{Cromartie:2019kug,2021arXiv210400880F}, see Fig.~\ref{fig:MR_laleos}.

\section{Bayesian methodology} To perform Bayesian model-selection among several EoSs, we need to compute the Bayesian {\em evidence} for each model combining astrophysical data from multiple messengers. The Bayesian methodology used in this work is primarily based on the following Refs.~\citep{Del_Pozzo_2013,Landry_2020PhRvD.101l3007L,Biswas_arXiv_2008.01582B}, which is described here.

For any two given EoSs, the relative {\em odds ratio} between them can be computed as
\begin{equation}
    \mathcal{O}_{j}^{i} = \frac{P(\mathrm{EoS}_{i}|d)}{P(\mathrm{EoS}_{j}|d)},
\end{equation}
where $d = (d_{\rm GW}, d_{\rm X-ray}, d_{\rm Radio})$ is the set of data from the three different 
types of astrophysical observations. Now using the Bayes' theorem we find,

\begin{equation}
    \underbrace{\mathcal{O}_{j}^{i}}_{\text{odds ratio} } = \underbrace{\prod_k \frac{P ({d_k} | \mathrm{EoS}_{i})}{P ({d_k} | \mathrm{EoS}_{j})}}_{\text{Bayes factor}} \times \underbrace{\frac{P(\mathrm{EoS}_{i})}{P(\mathrm{EoS}_{j})}}_{\text{ratio of priors}},
    \label{odds ratio}
\end{equation}
where we assume independence between different sets of data and $P(\mathrm{EoS}_{i,j})$ is the prior on the $\mathrm{EoS}_{i,j}$ before any measurement has taken place. Here we assume each model is equally likely and set the ratio of priors between two models to 1. Therefore, the main quantity of interest is the {\em Bayes factor} ($\mathcal{B}_{j}^{i}$) between a candidate $\mathrm{EoS}_{i}$ and $\mathrm{EoS}_{j}$. When $\mathcal{B}_{j}^{i}$ is substantially positive, it implies that the data prefers $\mathrm{EoS}_{i}$ over $\mathrm{EoS}_{j}$.

For GW observations, information about EoS parameters come from the masses $m_1, m_2$ of the two binary components and the corresponding tidal deformabilities $\Lambda_1, \Lambda_2$. In this case, 
\begin{align}
    P(d_{\mathrm{GW}}|\mathrm{EoS}) = \int^{M_{\mathrm{max} }}_{m_2}dm_1 \int^{m_1}_{M_{\mathrm{min} }} dm_2 P(m_1,m_2|\mathrm{EoS})   \nonumber \\
    \times P(d_{\mathrm{GW}} | m_1, m_2, \Lambda_1 (m_1,\mathrm{EoS}), \Lambda_2 (m_2,\mathrm{EoS})) \,,
    \label{eq:GW-evidence}
\end{align}
where $P(m_1,m_2|\rm{EoS})$ is the prior distribution over the component masses which should be informed by the NS population model. However, the choice of wrong population model starts to bias the results significantly only after $\sim 20\mbox{-}30$ observations~\citep{Agathos_2015,Wysocki-2020,Landry_2020PhRvD.101l3007L}. Therefore given the small number of detections at present, this can be fixed by a simple flat distribution over the masses,
\begin{equation}
    P(m|\rm{EoS}) = \left\{ \begin{matrix} \frac{1}{M_\mathrm{max} - M_\mathrm{min}} & \text{ iff } & M_\mathrm{min} \leq m \leq M_\mathrm{max}, \\ 0 & \text{ else, } & \end{matrix} \right.
\end{equation}
In our calculation we set $M_\mathrm{min} = 1M_{\odot}$ and $M_\mathrm{max}$ to the maximum mass for that particular EoS. Here the normalization factor on the NS mass prior is very important as it prefers the EoS with slightly larger $M_\mathrm{max}$ than the heaviest observed NS mass and disfavor EoS with much larger $M_\mathrm{max}$. For example, two EoSs with $M_\mathrm{max}=2.5 M_{\odot}$ and $M_\mathrm{max}=3 M_{\odot}$, will have mass prior probability $P(m|\rm{EoS})=2/3$ and $P(m|\rm{EoS})=1/2$ respectively if $M_\mathrm{min} \leq m \leq M_\mathrm{max}$. Though both EoSs support the heaviest NS mass measurement ($2.08 \pm 0.07 M_{\odot} $) equally well, EoS with $M_\mathrm{max}=3 M_{\odot}$ is less probable than EoS with $M_\mathrm{max}=2.5 M_{\odot}$. Similar approach has been employed in previous works as well~\citep{Miller:2019nzo,Raaijmakers:2019dks,Landry_2020PhRvD.101l3007L,Biswas:2021yge}. Alternatively, one can truncate the NS mass distribution to a largest population mass which is informed a formation channel (eg. supernova)---in that situation EoSs with $M_\mathrm{max}$ greater than the largest population mass will be assigned equal probability. Given our lack of knowledge on the upper limit of NS mass distribution, we choose to limit $M_\mathrm{max}$ based on EoS itself not the formation channel. A broader discussion on different choice of mass prior can be found in the appendix of Ref.~\citep{Legred:2021hdx}. However, if the masses in GW observation are expected to be smaller than the mass of the heaviest pulsar, then the upper limit on the mass prior can be chosen by the Likelihood's domain of support to reduce the computational time.

Equation~\ref{eq:GW-evidence} can be further simplified by fixing the GW chirp mass to its median value with not so much affecting the result~\citep{Raaijmakers:2019dks} given its high precision measurement. Then we will have one less parameter to integrate over as $m_2$ will be a deterministic function of $m_1$.

X-ray observations give the mass and radius measurements of NS. Therefore, the corresponding evidence takes the following form,
\begin{align}
    P(d_{\rm X-ray}|\mathrm{EoS}) = \int^{M_{\mathrm{max} }}_{M_{\mathrm{min} }} dm P(m|\mathrm{EoS}) \nonumber \\ \times
    P(d_{\rm X-ray} | m, R (m, \mathrm{EoS})) \,.
\end{align}
Similar to GW observation, here also the explicit prior normalization over the mass should be taken into account or can be chosen by the Likelihood's domain of support (if applicable). 

 Radio observations provide us with very accurate measurements of the NS mass. In this case, we need to marginalize over the observed mass taking into account its measurement uncertainties,
\begin{align}
    P(d_{\rm Radio}|\theta) = \int^{M_{\mathrm{max} }}_{M_{\mathrm{min} }}dm   P(m|\mathrm{EoS})
    P(d_{\rm Radio} | m) \,.
\end{align}
Here the prior normalization of mass must be taken into account as the observed mass measurement is close to the maximum mass predicted by the EoS. To compute these {\em evidences} we use the nested sampling algorithm implemented in {\tt Pymultinest}~\citep{Buchner:2014nha}
package. 

\section{Datasets} The likelihood distributions used in this work are modelled as follows:  (a). Mass and tidal deformability measurement from GW170817~\citep{Abbott:2018wiz} and GW190425~\citep{Abbott:2020uma} are modelled with an optimized multivariate Gaussian kernel density estimator (KDE) implemented in {\tt Statsmodels}~\citep{seabold2010statsmodels}. (b) Similarly mass and radius measurement of PSR J0030+0451~\citep{Riley:2019yda,Miller:2019cac} and PSR J0740+6620~\citep{Riley:2021pdl,Miller:2021qha} are also modelled with Gaussian KDE. Since the uncertainty in the mass-radius measurement of PSR J0740+6620 is larger for Ref.~\citep{Miller:2021qha} than Ref.~\citep{Riley:2021pdl} due to a conservative treatment of calibration error, we analyze both data separately and provide two {\em Bayes factor} values. (c) Mass measurement of PSR J0740+6620~\citep{Cromartie:2019kug,2021arXiv210400880F} should be modelled with a Gaussian likelihood of $2.08 M_{\odot}$ mean and $0.07 M_{\odot}$ $1 \sigma$ standard deviation. However we do not need to use this anymore, as the mass-radius measurement of PSR J0740+6620 already takes it into account. This is to note that further constraint on the properties of NS could be given by the joint detection of GW170817 and its electromagnetic counterparts~\citep{Bauswein_2017,Radice:2017lry,Coughlin:2018fis,Capano:2019eae,Breschi_2021}. However, this paper intentionally does not include that information as these constraints are rather indirect and need careful modeling of the counterparts. Similarly, constraints coming from heavy ion collision data~\citep{Margueron:2017eqc,Margueron:2018eob} are not included in this study. For example, recent measurement of neutron skin thickness of $\rm Pb^{208}$~\citep{Adhikari:2021phr} may suggest relatively stiff EoS around the nuclear saturation density, however present constraints are mostly dominated by the astrophysical observations~\citep{Essick:2021kjb,Biswas:2021yge}.  Nevertheless, in future it would be interesting to extend this analysis in that direction.

In the past, one common approach has been opted in many papers is to use the bound of radius and tidal deformability of $1.4 M_{\odot}$ NS to rule out EoSs of NS; rather than using the full distribution of the dataset. This can lead to a significant bias in the results~\citep{Miller:2019nzo}. Those bounds are also based on either a particular EoS parameterization or EoS insensitive relations~\citep{Maselli:2013mva,Yagi_2016,Yagi_2017,Chatziioannou_2018}. Different EoS parameterization leads to different bounds as they do not occupy the same prior volume. In contrast, the results obtained in this study do not depend on any EoS parameterization and therefore, these are also not subjected to any bias due to the model assumption. However, we are also using a specific set of EoSs which restrict us to provide accurate description of preferred EoS model. If the true model is among the candidates, then Bayesian model-selection method will select the correct one. But if all the models are false, then Bayesian model-selection will only select the least incorrect one.

\begin{figure*}[p]
    \centering
\begin{tabular}{cc}
    \includegraphics[width=0.45\textwidth]{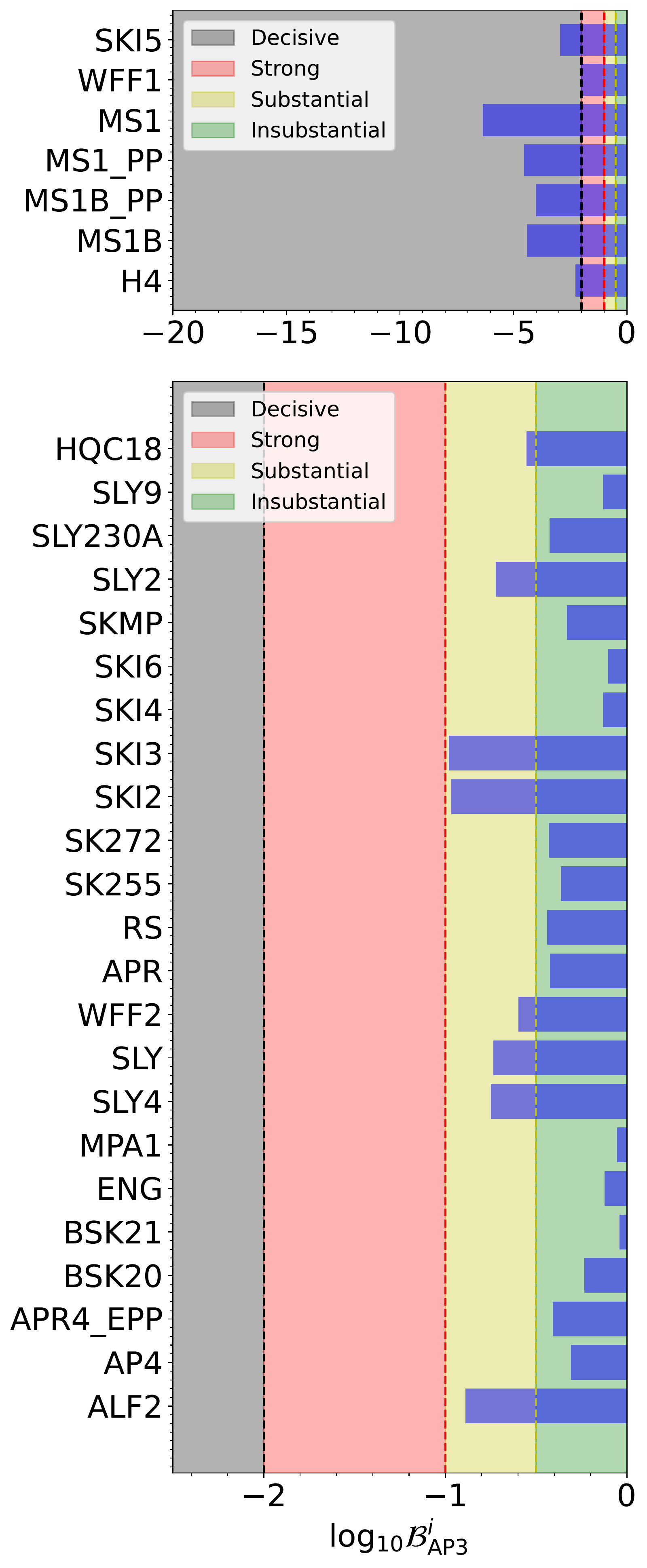}&
    \includegraphics[width=0.45\textwidth]{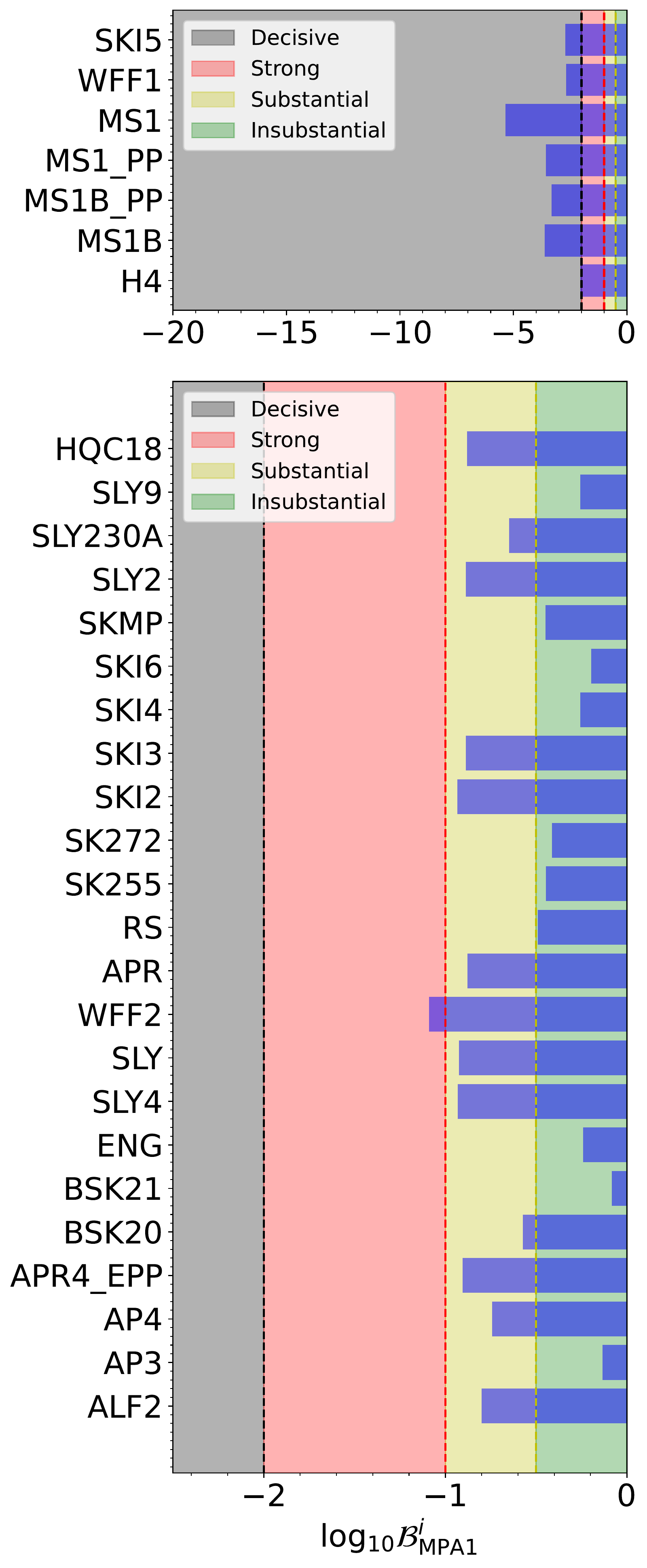}\\
\end{tabular}

   \caption{{\em Bayes factor} for different EoS models are plotted with respect to the most preferred EoS AP3 (or MPA1). Following the interpretation of Kass and Raftery~\citep{Bayes-factor}, we have divided the {\em Bayes factor} values into four different regions: (a) $\log_{10}\mathcal{B}_{\rm AP3/MPA1}^{i} \leq -2$: decisive evidence against AP3/MPA1, (b) $-2 < \log_{10}\mathcal{B}_{\rm AP3/MPA1}^{i} \leq -1$: strong evidence against AP3/MPA1, (c) $-1 < \log_{10}\mathcal{B}_{\rm AP3/MPA1}^{i} \leq -1/2$: substantial evidence against AP3/MPA1, and (d) $\log_{10}\mathcal{B}_{\rm AP3/MPA1}^{i} \geq -1/2$: insubstantial evidence against AP3/MPA1. }
    \label{fig:bayes-factor-comparison} 

\end{figure*}

\section{Results} 
\begin{figure}[ht!]
    \centering
    \includegraphics[width=0.45\textwidth]{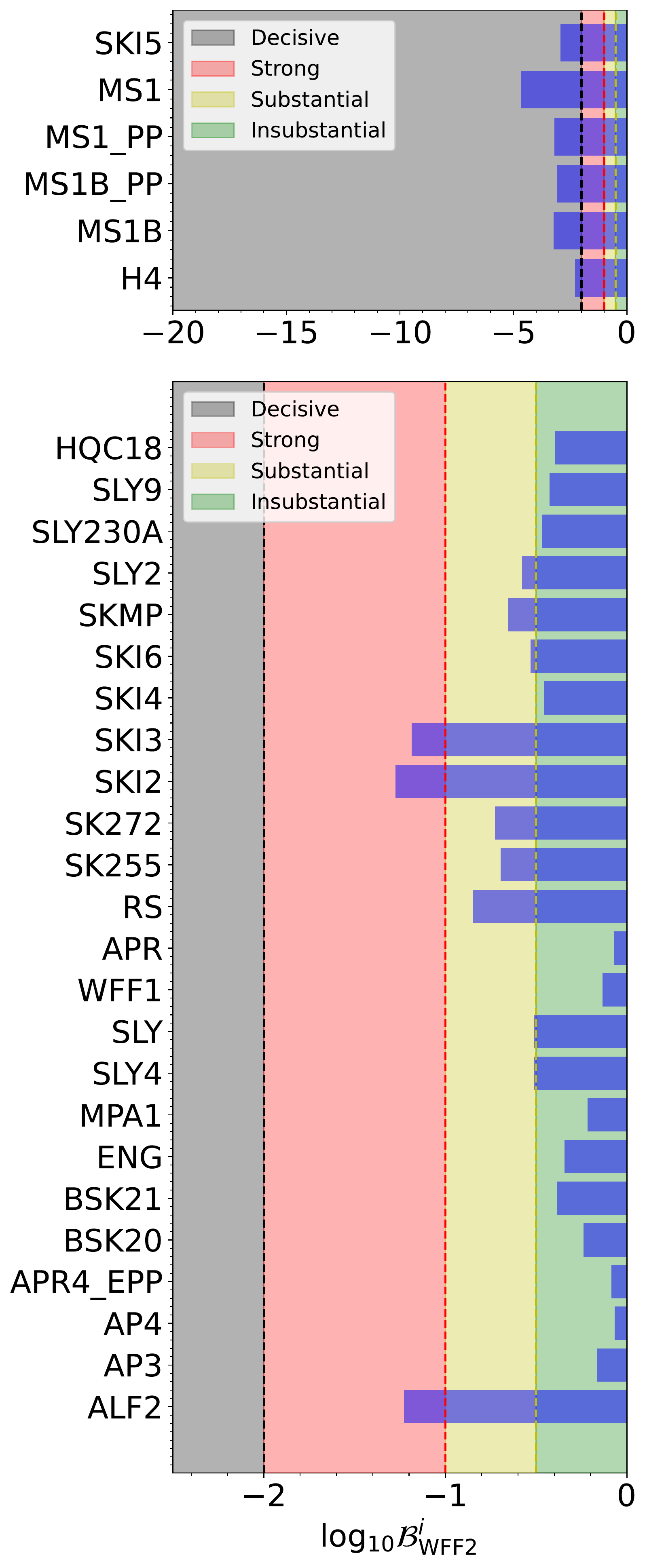}
   \caption{This Fig. is same as Fig.~\ref{fig:bayes-factor-comparison} but only combining two GW observations and the revised mass-measurement of PSR J0740+6620 by Ref.~\citep{2021arXiv210400880F}. In this case, WFF2 turns out to be most preferred EoS.}
    \label{fig:bayes-factor-comparison-GW}
\end{figure}

In figure~\ref{fig:bayes-factor-comparison}, {\em Bayes factor} of different EoSs are plotted using multi-messenger observations with respect to the most probable EoS, for which we find the Bayesian {\em evidence} to be maximum. In the left panel results are obtained using the data from Ref.~\citep{Riley:2019yda,Riley:2021pdl} for which AP3 turns out to be the most preferred model. In the right panel the data from Ref.~\citep{Miller:2019cac,Miller:2021qha} are used and in this case, MPA1 is the most preferred EoS. MPA1 EoS has a larger value of $R_{1.4}$, $\Lambda_{1.4}$, and $M_{\rm max}$ compare to AP3 EoS. Therefore, it is clear the data from Ref.~\citep{Miller:2019cac,Miller:2021qha} prefer stiffer EoS compare to the data from Ref.~\citep{Riley:2019yda,Riley:2021pdl}. In this study, we follow the interpretation of Kass and Raftery~\citep{Bayes-factor} and decisively exclude the EoSs for which $\log_{10}\mathcal{B}_{\rm AP3}^{i} \leq -2$. This region is shown using black shade in the plot. We find SKI5, WFF1, MS1, $\rm MS1\_{PP}$, $\rm MS1B\_{PP}$, MS1B, and H4 are ruled out for both type of datasets. All of these EoSs except WFF1 are rather stiff EoSs and predict large values of radius and tidal deformability for the NS.  This is broadly consistent with GW170817 observations as it mainly favors soft EoS~\citep{TheLIGOScientific:2017qsa,Abbott:2018exr}. However WFF1 which is the softest EoS considered in this study, is also now decisively ruled out by the multi-messenger observations. In fact WFF1 was found to be the one of the most preferred EoS by the previous studies~\citep{LIGOScientific:2019eut,Ghosh:2021eqv,Pacilio:2021jmq} based on GW observation only (see also Fig.~\ref{fig:bayes-factor-comparison-GW}). This demonstrates the true power of multi-messenger observations. Now not only stiff EoSs but also extreme soft EoSs are ruled out.

The region between $-2 < \log_{10}\mathcal{B}_{\rm AP3/MPA1}^{i} \leq -1$ is shown in red shade. Only WFF2 falls in this region while using the data from Ref.~\citep{Miller:2019cac,Miller:2021qha} and they have strong evidence against MPA1 according to the interpretation of Kass and Raftery. WFF2 is a relatively softer EoS with the value of $R_{1.4} = 11.16$ km and $\Lambda_{1.4}=232$. 

$\log_{10}\mathcal{B}_{\rm AP3/MPA1}^{i} > -1$ are statistically insignificant. However it is still divided into two regions: $-1 < \log_{10}\mathcal{B}_{\rm AP3/MPA1}^{i} \leq -1/2$ is referred to as substantial evidence against AP3 (shown in yellow band) and $\log_{10}\mathcal{B}_{\rm AP3/MPA1}^{i} \geq -1/2$ is referred to as insubstantial (shown in green band), {\em which is not worth more than a bare mention}. Most of our candidate EoSs lie in these two regions; 8(13) of them have substantial evidence and 15(10) of them have  insubstantial evidence against AP3(MPA1). The status of each EoS model is also summarized in the last column of Table~\ref{tab:eos_table} using the data from~\citep{Riley:2019yda,Riley:2021pdl}.

Despite a slightly different radius measurement of PSR J0740+6620 between~\citep{Riley:2021pdl} and~\citep{Miller:2021qha}, we find broadly consistent Bayesian evidence using both datasets. From the previous studies~\citep{Biswas:2021yge,Legred:2021hdx}, it is expected of a maximum $\sim 0.4$ km difference in $R_{1.4}$ towards the stiff EoSs when the data from
from~\citep{Miller:2021qha} is added instead of~\citep{Riley:2021pdl}. In our study, we also find the similar trend. There is a $0.4$ km difference in $R_{1.4}$ between the two most preferred EoSs. One way to remove this systematic error, is to integrate over both datasets using Bayesian framework. Naively one can take equal number of samples from the posterior using both datasets and combined them into a single posterior. However as discussed in~\citep{Ashton:2019leq} on the systematic error due to modeling
uncertainty in waveform for GW signal, giving equal-weighted probability on the both datasets might not be correct way as they corresponds to different Bayesian evidence i.e, one being more likely compare to the other. It is better to weight the samples by their corresponding posterior evidence. Since this paper only focuses on the Bayesian evidence and does not deal with the posterior of any inferred properties, we leave this discussion for a future work~\citep{BD_universalilty}.

\begin{longtable}[ht]{|p{2.1cm}|p{.8cm}|p{.8cm}|p{.8cm}|p{2.1cm}|}

\hline
\toprule
     EoS & $R_{1.4}$ [km] & $\Lambda_{1.4}$ & $M_{\rm max} \newline  (M_{\odot})$ &         Evidence \newline against AP3 \\
\midrule
\hline
    ALF2~\citep{Alf2} &     13.19 &       759 &          2.09 &    Substantial \\
     AP3~\citep{APR4_EFP} &     12.10 &       393 &          2.39 & Most preferred \\
     AP4~\citep{APR4_EFP} &     11.43 &       263 &          2.21 &  Insubstantial \\
APR4\_EPP~\citep{Endrizzi_2016,APR4_EFP} &     11.32 &       248 &          2.16 &  Insubstantial \\
   BSK20~\citep{Goriely_2010,Pearson_2011} &     11.76 &       324 &          2.17 &  Insubstantial \\
   BSK21~\citep{Goriely_2010,Pearson_2011} &     12.60 &       530 &          2.28 &  Insubstantial \\
     ENG~\citep{Engvik_1994} &     12.12 &       414 &          2.25 &  Insubstantial \\
      H4~\citep{H4} &     13.69 &       897 &          2.03 &       Decisive \\
    MPA1~\citep{MPA1} &     12.50 &       513 &          2.47 &  Insubstantial \\
    MS1B~\citep{MS_PP} &     14.68 &      1409 &          2.80 &       Decisive \\
 MS1B\_PP~\citep{MS_PP} &     14.53 &      1225 &          2.75 &       Decisive \\
  MS1\_PP~\citep{MS_PP} &     14.93 &      1380 &          2.75 &       Decisive \\
     MS1~\citep{MS_PP} &     15.07 &      1622 &          2.80 &       Decisive \\
    SLY4~\citep{Douchin:2001sv} &     11.78 &       313 &          2.05 &    Substantial \\
     SLY~\citep{Douchin:2001sv} &     11.78 &       313 &          2.05 &    Substantial \\
    WFF1~\citep{wff1} &     10.42 &       153 &          2.14 &       Decisive \\
    WFF2~\citep{wff1} &     11.16 &       232 &          2.20 &    Substantial \\
     APR~\citep{APR4_EFP} &     11.35 &       249 &          2.19 &  Insubstantial \\
      RS~\citep{SLy2_1,SLy2_2,RS_3} &     12.92 &       591 &          2.12 &    Insubstantial \\
   SK255~\citep{SLy2_1,SLy2_2,SK255_3} &     13.14 &       586 &          2.14 &    Insubstantial \\
   SK272~\citep{SLy2_1,SLy2_2,SK255_3} &     13.30 &       642 &          2.23 &    Insubstantial \\
    SKI2~\citep{SLy2_1,SLy2_2,REINHARD1995467} &     13.47 &       770 &          2.16 &         Substantial \\
    SKI3~\citep{SLy2_1,SLy2_2,REINHARD1995467} &     13.54 &       785 &          2.24 &         Substantial \\
    SKI4~\citep{SLy2_1,SLy2_2,REINHARD1995467} &     12.37 &       469 &          2.17 &  Insubstantial \\
    SKI5~\citep{SLy2_1,SLy2_2,REINHARD1995467} &     14.07 &      1009 &          2.24 &       Decisive \\
    SKI6~\citep{SLy2_1,SLy2_2,REINHARD1995467} &     12.48 &       492 &          2.19 &  Insubstantial \\
    SKMP~\citep{SLy2_1,SLy2_2,SKMP_3} &     12.49 &       478 &          2.11 &    Insubstantial \\
    SLY2~\citep{SLy2_1,SLy2_2} &     11.78 &       310 &          2.05 &    Substantial \\
 SLY230A~\citep{SLy2_1,SLy2_2,SLy230A_3} &     11.83 &       330 &          2.10 &  Insubstantial \\
    SLY9~\citep{SLy2_1,SLy2_2} &     12.46 &       450 &          2.16 &  Insubstantial \\
   HQC18~\citep{HQC18} &     11.49 &       257 &          2.05 &    Substantial \\
\bottomrule
\hline
\caption{List of all the 31 EoSs with corresponding radius and tidal deformability of a $1.4 M_{\odot}$ NS are provided here. At the last column the status of each EoS is mentioned using the multi-messenger observations of NSs (for X-ray, the data from~\citep{Riley:2019yda,Riley:2021pdl} are used here) based on their {\em Bayes factor} value with respect to the most preferred EoS (AP3).}

\label{tab:eos_table}

\end{longtable}

\section{Summary} In summary, we have performed a Bayesian model-selection study over a wide-ranging EoS model using multi-messenger observations of NS. We find EoSs which predict larger radius ($R_{1.4} \geq 13.69$ km) and tidal deformability ($\Lambda_{1.4} \geq 897$ ) are decisively ruled out.  As well as WFF1 EoS which predicts a lower radius ($R_{1.4} = 10.42$ km) and tidal deformability ($\Lambda_{1.4} = 153$ ) is also decisively ruled out. Therefore, EoS of NS cannot be either very stiff or soft. Ironically, the range of $R_{1.4}$ from this analysis is very consistent with the prediction ($11.5-13.6$ km) based on the nuclear physics data by~\citep{Li:2005sr}. The result obtained in this work gives the current status of various EoS models and can be used as a benchmark while making new EoS models or choosing an EoS model to perform any numerical simulation of NS. Any future measurement of NS properties from GW and electromagnetic observations can be easily combined using the methodology developed in this paper. 
\section*{Acknowledgements} I am extremely thankful to Nikolaos Stergioulas for reading this manuscript carefully and making several useful suggestions. I also thank Prasanta Char, Rana Nandi, Wolfgang Kastaun, and Sukanta Bose for their valuable suggestions to improve this manuscript. I gratefully acknowledge the use of high performance super-computing cluster Pegasus at IUCAA for this work. I acknowledge support from the Knut and Alice Wallenberg Foundation 
under grant Dnr. KAW 2019.0112. This material is based upon work supported by NSF’s LIGO Laboratory which is a major facility fully funded by the National Science Foundation.

\bibliography{mybiblio}
\label{lastpage}
\end{document}